\documentclass{article} 
\usepackage{amsmath,amsfonts,amssymb,multicol}
\usepackage{leftidx}
\usepackage{graphicx,caption,subcaption}
\usepackage[usenames]{xcolor}
\usepackage{graphicx}
\usepackage{lipsum}
\usepackage{euscript,amssymb}
\usepackage{epsfig}
\usepackage{fancyhdr}
\usepackage{esint}
\usepackage{color}

{\definecolor{RED}{rgb}{1,0,0}
\definecolor{GREEN}{rgb}{0,1,0}
\definecolor{BLUE}{rgb}{0,0,1}

\def\be{\begin{equation}}
\def\ee{\end{equation}}
\def\bea{\begin{eqnarray}}
\def\eea{\end{eqnarray}}
\usepackage{amsfonts}

\usepackage{amsmath}

\usepackage{amssymb}

\usepackage{fancyhdr}
\def\nn{\nonumber\\}
\def\be{\begin{equation}}
\def\ee{\end{equation}}
\def\bea{\begin{eqnarray}}
\def\eea{\end{eqnarray}}

\def\bwt{\begin{widetext}}
\def\ewt{\end{widetext}}

\usepackage{esint}
\usepackage[unicode=true, pdfusetitle,
 bookmarks=true,bookmarksnumbered=false,bookmarksopen=false,
 breaklinks=false,pdfborder={0 0 1},backref=false,colorlinks=false]
 {hyperref}
\def\T{T}

\newcommand{\f}[2]{\frac{#1}{#2}}
\begin{document}
\title{Dynamic wormhole solutions in  Einstein-Cartan gravity}
\author{Mohammad Reza Mehdizadeh$^{1\!\!}$ \footnote{mehdizadeh.mr@uk.ac.ir}\,\, and\, Amir Hadi Ziaie$^2$\footnote{ah.ziaie@gmail.com}
	\\\\$\leftidx{^1}{{\rm Department~of~ Physics,~ Shahid~ Bahonar~ University, P.~ O.~ Box~ 76175, Kerman, Iran}}$ 
	\\\\
	$\leftidx{^2}{{\rm Department~of~ Physics,~ Kahnooj~Branch,~Islamic~Azad~University, 7881916145~Kerman,~Iran}}$}
\date{\today}
\maketitle
\begin{abstract}
In the present work we investigate evolving wormhole configurations described by a constant redshift function in Einstein-Cartan theory ({{\sf ECT}}). The matter content consists of a Weyssenhoff fluid along with an anisotropic matter which together generalize the anisotropic energy momentum tensor in general relativity ({{\sf GR}}) in order to include the effects of intrinsic angular momentum (spin) of particles. Using a generalized Friedmann-Robertson-Walker ({\sf FRW}) spacetime, we derive analytical evolving wormhole geometries by assuming a particular equation of state ({{\sf EoS}}) for energy density and pressure profiles. We introduce exact asymptotically flat and  anti-de Sitter spacetimes that admit traversable wormholes and respect energy conditions throughout the spacetime. The rate of expansion of these evolving wormholes is determined only by the Friedmann equation in the presence of spin effects.
\end{abstract}
\section{Introduction}
Wormholes are theoretical passages in the spacetime topology that could create handles or tunnels which link two parallel universes or widely separated regions of the same Universe. The search for exact spacetimes admitting wormhole solutions in {\sf GR} has attracted a great deal of interest in theoretical physics within different field of studies. Much work has been done over the past decades in order to explain physics as pure geometry namely within the ancient Einstein-Rosen bridge model of a particle \cite{ERBridge}, see also \cite{BrillLindquist}. The concept of wormhole was invented in the late 1950\rq{}s within the pioneering articles of Misner and Wheeler \cite{misner-wheeler} and Wheeler \cite{Wheelerworm}, in order to provide a mechanism for having \lq\lq{}charge without charge\rq\rq{}. The electric charge was claimed to emerge as a manifestation of the topology of a space which in essence resembled a sheet with a handle. The name of such an object was proposed as a \lq\lq{}wormhole\rq\rq{}. Despite its elegance and simplicity, the interest in Misner-Wheeler wormhole declined over the years firstly due to the rather ambitious nature of the project which unfortunately had little connection with the real world or support from Earth based experiments (see e.g., and reference therein \cite{kar-sahdev}).
The study of Lorentzian wormholes in the context of {\sf GR} was triggered by the remarkable paper of Morris and Thorne in 1988 \cite{mt} where, they introduced a static spherically symmetric metric and discussed the required conditions for physically meaningful traversable wormholes. In {\sf GR}, the fundamental faring-out condition of wormhole throat leads to the violation of null energy condition ({\sf NEC}). The matter distribution responsible for {\sf NEC} violation is the so called \lq\lq{}{\textrm exotic matter}\rq\rq{} \cite{khu}, by the virtue of which, traversable wormhole geometries have been obtained e.g., with the help of phantom energy distribution \cite{phantworm}. This type of matter, though exotic in the Earth based laboratory context, is of observational interest in cosmological settings \cite{phan-dark-sce}. Phantom energy possesses peculiar features, namely, a divergent cosmic energy density at a finite time\cite {phant1}, prediction of existence of a new long range force \cite{phant2}, and the appearance of a negative entropy and negative temperature~\cite{phant3}.
\par
One of the most important challenges in wormhole scenarios is the establishment of standard energy conditions. In this regard, various methods have been proposed  in the literature that deal with the issue of energy conditions within wormhole settings. Moreover, Visser and Poisson have studied the construction of thin-shell wormholes where the supporting matter is concentrated on the wormhole\rq{}s throat \cite{pv}. Fortunately, in the context of modified theories of gravity, the presence of higher order terms in curvature would allow for building thin-shell wormholes supported by ordinary matter \cite{thi}. Recently, a large amount of work has been devoted to build and study wormhole solutions within the framework of modified gravity theories among which we can quote: wormhole solutions in Brans-Dicke theory \cite{bd}, $f({\sf R})$ gravity \cite{fr}, Born-Infeld theory~\cite{bf}, Einstein-Gauss-Bonnet theory~\cite{gmfl}, Kaluza-Klein gravity \cite{kl} and scalar-tensor gravity~\cite{kash}.
\par
Though the exotic energy-momentum is required to support wormhole configurations~\cite{exoworm}, it has been noted that the possibility of evolving (time-dependent) wormholes may help to modify this situation~\cite{tddynworm}. Work along this line has been done in dynamical wormhole geometries which satisfy the energy conditions during a time period \cite{kst}. An interesting scenario is that the expansion of the Universe could increase the size of the static wormholes by a factor which is proportional to the scale factor of the Universe. In this regard, a pioneering work related to dynamical wormholes was done independently by Hochberg and Visser \cite{hves1} and Hayward \cite{sh1}, however, the supporting matter violates the {\sf NEC}. Evolving wormholes in a cosmological background have been also studied in \cite{kl4}. Cataldo et al. studied the $(N +1)$ dimensional evolving wormholes supported by a polytropic {\sf EoS}~\cite{maf}. Maeda, Harada, and Carr have shown another class of dynamical wormholes (cosmological wormholes) which mimic an asymptotically Friedmann Universe with a big-bang singularity at the beginning~\cite{mda1}. The resulted wormhole spacetimes contain a perfect fluid and admit a homothetic Killing vector which requires an {\sf EoS} to be linear which indeed makes the cosmic expansion to be accelerating for an appropriate {\sf EoS} parameter. However, for these class of solutions the numerical solution exhibits a singular hypersurface which violates the {\sf NEC}. In recent years, some research works have been done that deal with dynamic wormhole spacetimes supported by two fluids \cite{gm3} and evolving wormholes sustained by a single inhomogeneous and anisotropic fluid for which a generalized {\sf EoS} is imposed~\cite{rbo}.
\par
The {\sf ECT} is a gravitational theory which was put forward by the desire to provide a simple description of the effects of spin on gravitational interactions \cite{ECT3,ECT4}. This can be achieved by taking as a model of spacetime a four-dimensional differential manifold endowed with a metric tensor and a linear connection which is asymmetric. The torsion tensor refers to the anti-symmetric part of the connection which physically can be interpreted as caused by the spin of fermionic matter fields. Hence in {\sf ECT}, both mass and spin, which are intrinsic and fundamental properties of matter would influence the structure of spacetime.
\par
While {\sf GR} has been a successful theory in describing the gravitational phenomena, this theory admits spacetime singularities both in the cosmological and astrophysical settings \cite{hawellis}. These are spacetime events where the densities as well as curvatures grow boundlessly and the classical framework of the theory breaks down. It is therefore well motivated to search for alternative theories of {\sf GR} whose geometrical attributes may provide nontrivial settings in order to study the gravitational interactions. In this regard, one of the advantages of introducing torsion is to modify the present standard cosmology based on usual {\sf GR} by means of the spin of matter. On the other hand, the standard model of cosmology is built upon the homogeneity and isotropy of the Universe on large scales while being inhomogeneous on small scales. This model can be extended to inhomogeneous spherically symmetric spacetimes (which merge smoothly to the cosmological background) by assuming that the radial pressure and energy density obey a linear equation of state ({\sf EoS}), i.e., $p_r=w\rho$. An interesting scenario is due to the fact that the expansion of the Universe could increase the size of the static wormholes by a factor which is proportional to the scale factor of the Universe. Using a linear {\sf EoS}, wormhole solutions have been obtained in {\sf GR} and their physical properties have been discussed in \cite{sulpa}. Cosmological settings in {\sf ECT} has been also investigated where it has been shown that spacetime torsion may provide a setting in which the big bang singularity is replaced by a non-singular state of minimum but finite radius \cite{spin-bounce}. Moreover, torsion has been employed to study the spin effects in the early Universe~\cite{Gas},\cite{earlyunimex}, inflationary models \cite{torearlyinflation}, emergent Universe scenario \cite{hhhd}, gravitational collapse \cite{colltorspin}, higher dimensional gravity theories \cite{higherordertorsion} and black hole physics \cite{bhphystor}. Recently, the possibility of existence of static traversable wormholes in the context of {\sf ECT}, without resorting to an exotic matter, has been studied in \cite{Broni-twoscalarfield}. Taking the matter sources as two noninteracting scalar fields (one is minimally and the other is nonminimally coupled to gravity) with nonzero potentials, exact static, spherically symmetric wormhole solutions with flat or AdS asymptotic behavior has been obtained. These kind of wormholes satisfy the {\sf NEC} and weak energy condition ({\sf WEC}) with arbitrary throat radius. More interestingly, exact wormhole spacetimes with sources in the form of a nonminimally coupled
non-phantom scalar field and an electromagnetic field have been found in \cite{Broniprd2016}. The solutions describe different asymptotic behavior and symmetric properties and a minimum value for the throat radius has been obtained subject to satisfaction of {\sf NEC} and {\sf WEC}. Work along this line has been also performed in \cite{mehdihadi} where wormhole structures and the energy conditions supporting them have been studied in the framework of {\sf ECT}. Introducing the supporting material for wormhole geometry as a Weyssenhoff spinning fluid along with an anisotropic energy momentum tensor ({\sf EMT}) for matter fields, exact asymptotically flat and anti-de-Sitter spacetimes were obtained which admit traversable wormholes and respect energy conditions.
\par

\par
Motivated by the above considerations, we search for the dynamical wormhole solutions in a cosmological background in {\sf ECT}. We are interested to study spherically symmetric dynamical wormhole solutions in a cosmological background which are supported by an anisotropic spinning fluid. The matter content supporting the wormhole geometry includes the {\sf EMT} of a spinning fluid together with an anisotropic {\sf EMT} for the ordinary matter distribution. As we shall see, two classes of traversable wormhole solutions satisfying {\sf WEC} can be found for suitable values of the {\sf EoS} parameter.

\par
This paper is organized as follows: In section \ref{EC} we give a brief review on {\sf ECT} together with finding the gravitational field equations. Introducing a spin fluid as the source of spacetime torsion, we rewrite the field equations for an anisotropic source and present the resulted differential equations governing the wormhole configuration. Taking an {\sf EoS} for the radial and tangential pressures and energy density, we deal with evolving wormhole solutions in the context of {\sf ECT} in section \ref{WHS}. Two classes of solutions are found as dynamical wormhole solutions with zero tidal force, presented in subsections \ref{3.2ew} and \ref{3.3ew}. Our conclusion is drawn in section \ref{concluding}.
\section{Field equations in Einstein-Cartan theory}\label{EC}
In the framework of {\sf GR}, the gravitational field is described by a symmetric rank two tensor, i.e., the metric tensor which is defined on a four-dimensional spactime manifold. The Einstein-Hilbert action then provides the dynamics of this tensor field via the {\sf GR} field equations. However, it is possible to generalize the {\sf GR} action through defining much more invariants from the spacetime torsion and curvature tensors. The {\sf ECT} allows us to proceed in this sense and find the simplest and most natural generalization of {\sf GR}, for which the action integral is given by
\bea S&=&\int  d^4x\sqrt{-{\sf g}}
\left\{\f{-1}{2\kappa^2}\left(\hat{{\sf R}}+2\Lambda\right)+\mathcal{L}_m\right\}\nonumber\\
&=&\int d^4x \sqrt{-{\sf g}}\bigg\{\f{-1}{2\kappa^2}\bigg[{\sf R}(\{\})+
{\sf K}^{\alpha}\!\!~_{\rho\lambda}{\sf K}^{\rho\lambda}\!\!~_{\alpha}-{\sf K}^{\alpha}\!\!~_{\rho\alpha}{\sf K}^{\rho\lambda}\!\!~_{\lambda}\bigg]+{\mathcal L}_m\bigg\}, \label{action}
\eea
where $\kappa^2=8\pi G$ is the gravitational coupling constant, $\hat{{\sf R}}$ is the Ricci scalar constructed out of a general asymmetric connection $\hat{\Gamma}^{\alpha}_{~\mu\nu}$ and can be expressed in terms of the independent background fields, i.e., the metric field ${\sf g}_{\mu\nu}$ and the connection. The quantity ${\sf K}^{\mu}_{\,\,\,\nu\alpha}$ is the contorsion tensor defined as
\be\label{contortion}
{\sf K}^{\mu}_{~\alpha\beta}={\sf Q}^{\mu}_{~\alpha\beta}+{\sf Q}_{\alpha\beta}^{~~\,\mu}+
{\sf Q}_{\beta\alpha}^{~~\,\mu},
\ee
where the spacetime torsion ${\sf Q}^{\alpha}_{~\mu\nu}$ is geometrically defined as the antisymmetric part of the connection\be\label{TT}
{\sf Q}^{\mu}_{~\alpha\beta}=\f{1}{2}\left[\hat{\Gamma}^{\mu}_{~\alpha\beta}-\hat{\Gamma}^{\mu}_{~\beta\alpha}\right].
\ee
The Lagrangian of the matter fields is introduced as $\mathcal L_m$. Extremizing action (\ref{action}) with respect to contorsion gives the Cartan field equation as
\be\label{FEEC}
{\sf Q}^{\alpha}_{~\mu\beta}-\delta^{\alpha}_{\,\beta}{\sf Q}^{\gamma}_{~\,\mu\gamma}+\delta^{\alpha}_{\,\mu}{\sf Q}^{\gamma}_{~\,\beta\gamma}=-\f{1}{2}\kappa\Sigma_{\mu\beta}^{~~\alpha},
\ee
or equivalently
\be\label{torspin}
{\sf Q}^{\alpha}_{~\mu\beta}=-\f{\kappa}{2}\left[\Sigma_{\mu\beta}^{\,\,\,\,\alpha}+\f{1}{2}\delta^{\alpha}_{\mu}\Sigma_{\beta\rho}^{\,\,\,\,\,\rho}-\f{1}{2}\delta^{\alpha}_{\beta}\Sigma_{\mu\rho}^{\,\,\,\,\,\rho}\right],
\ee
where $\Sigma^{\mu\alpha\beta}=2\left(\delta\mathcal L_m/\delta {\sf K}_{\mu\alpha\beta}\right)/\sqrt{-{\sf g}}$ is defined as the spin tensor of matter \cite{ECT3}. We note that the equation governing the torsion tensor is of algebraic type, thus, the torsion is not allowed to propagate outside the matter distribution as a torsion wave or through any interaction of non-vanishing range \cite{ECT3} and therefore is only nonzero inside the matter source. Varying action (\ref{action}) with respect to the metric gives the Einstein-Cartan field equation \cite{ECT3,Venzo}
\bea\label{ecfieldeq}
{\sf G}_{\mu\beta}\left(\{\}\right)-\Lambda g_{\mu\nu}=\kappa^2\left({{\sf T}}_{\mu\beta}+\theta_{\mu\beta}\right),
\eea
where
\bea\label{correctionterms}
\theta_{\mu\nu}&=&\f{1}{\kappa^2}\Bigg[4{\sf Q}^{\eta}_{\,\,\mu\eta}{\sf Q}^{\beta}_{\,\,\nu\beta}-\left({\sf Q}^{\rho}_{\,\,\,\mu\epsilon}+2{\sf Q}_{(\mu\epsilon)}^{\,\,\,\,\,\,\,\rho}\right)\left({\sf Q}^{\epsilon}_{\,\,\nu\rho}+2{\sf Q}_{(\nu\rho)}^{\,\,\,\,\,\,\,\,\epsilon}\right)+\f{1}{2}g_{\mu\nu}\left({\sf Q}^{\rho\sigma\epsilon}+2{\sf Q}^{(\sigma\epsilon)\rho}\right)\left({\sf Q}_{\epsilon\sigma\rho}+2{\sf Q}_{(\sigma\rho)\epsilon}\right)\nonumber\\&-&2g_{\mu\nu}{\sf Q}^{\rho\sigma}_{\,\,\,\,\rho}{\sf Q}^{\sigma}_{\,\,\,\epsilon\sigma}
\Bigg].\eea
or equivalently
\bea\label{SSCI}
\theta_{\mu\beta}&=&\f{1}{2}\kappa^2\bigg[\Sigma_{\mu\alpha}^{~~~\alpha}\Sigma_{\beta\gamma}^{~~~\!\!\gamma}-\Sigma_{\mu}^{~\alpha\gamma}\Sigma_{\beta\gamma\alpha}-\Sigma_{\mu}^{~\alpha\gamma}\Sigma_{\beta\alpha\gamma}\nn&+&\f{1}{2}\Sigma^{\alpha\gamma}_{~~~\mu}\Sigma_{\alpha\gamma\beta}+\f{1}{4}g_{\mu\beta}\left(2\Sigma_{\alpha\gamma\epsilon}\Sigma^{\alpha\epsilon\gamma}
-2\Sigma_{\alpha~\gamma}^{~\gamma}\Sigma^{\alpha\epsilon}_{~~~\epsilon}
+\Sigma^{\alpha\gamma\epsilon}\Sigma_{\alpha\gamma\epsilon}\right)\bigg],
\eea
where we have used expression (\ref{torspin}) in order to substitute for the torsion tensor, $()$ denotes symmetrization and ${{\sf T}}_{\mu\beta}$ being the dynamical {\sf EMT} represented by ${{\sf T}}_{\mu\beta}=2\left(\delta\mathcal L_m/\delta {\sf g}^{\mu\beta}\right)/\sqrt{-{\sf g}}$~\cite{spfieldspop}.
Next we proceed to obtain exact solutions exhibiting wormhole geometries in the presence of a spinning fluid. Such a fluid can be described by the so called Weyssenhoff fluid considered as a continuous macroscopic medium whose microscopic elements are composed of fermionic particles with intrinsic angular momentum. This model which generalizes the {\sf EMT} of ordinary matter in {\sf GR} to include non-vanishing spin was first studied by Weyssenhoff and Raabe \cite{W1947} and extended by Obukhov and Korotky in order to construct cosmological models based on the EC theory \cite{KCQG1987}. In order to consider wormhole solutions in the framework of {\sf ECT}, we utilize a classical description of spin as postulated by Weyssenhoff given by~\cite{W1947},\cite{KCQG1987},
\be\label{FC}
\Sigma_{\mu\nu}^{~~\alpha}={\sf S}_{\mu\nu}{\rm u}^{\alpha},~~~~~~~~{\sf S}_{\mu\nu}{\rm u}^{\mu}=0,
\ee 
where ${\rm u}^{\alpha}$ is the four-velocity of the fluid element and ${\sf S}_{\mu\nu}=-{\sf S}_{\nu\mu}$ is a second-rank antisymmetric tensor defined as the spin density tensor. The spatial components of spin density tensor include the 3-vector $({\sf S}^{23},{\sf S}^{13},{\sf S}^{12})$ which coincides in the rest frame with the spatial spin density of the matter element. The rest of spacetime components $({\sf S}^{01}, {\sf S}^{02},{\sf S}^{03})$ are assumed to be zero in the rest frame of fluid element, which can be covariantly formulated as a constraint given in the second part of (\ref{FC}). This constraint on the spin density tensor is usually called the Frenkel condition which requires the intrinsic spin of matter to be spacelike in the rest frame of the fluid\footnote{The Weyssenhoff spin fluid has been also described by means of applying the Papapetrou-Nomura-Shirafuji-Hayashi method of multiple expansion in the Riemann-Cartan spacetime \cite{PNSHH1951} to the conservation law for the spin density (which results from the Bianchi identities in the EC gravity \cite{consref,LordTen}) in the point-particle approximation.}. The dynamical {\sf EMT} can be decomposed into the usual perfect fluid part, ${\sf T}^{\sf Pf}_{\mu\beta}$ and an intrinsic spin part ${\sf T}^{\sf s}_{\mu\beta}$ which can be written explicitly as \cite{W1947,Gas}
\bea\label{sfss}
{{\sf T}}_{\mu\beta}&=&{\sf T}^{\sf Pf}_{\mu\beta}+{\sf T}^{\sf s}_{\mu\beta}=\left\{(\rho+p_{t}){\rm u}_\alpha {\rm u}_\beta+p_{t}g_{\alpha\beta}+(p_r-p_t){\rm v}_\alpha {\rm v}_\beta\right\}\nonumber\\&+&{\rm u}_{(\alpha}{\sf S}_{\beta)}^{\,\,\,\mu}{\rm u}^{\nu}{{\sf K}}^{\rho}_{\,\,\mu\nu}{\rm u}_{\rho}+{\rm u}^{\rho}{{\sf K}}^{\mu}_{\,\,\rho\sigma}{\rm u}^{\sigma}{\rm u}_{(\alpha}{\sf S}_{\beta)\mu}-\f{1}{2}{\rm u}_{(\alpha}{{\sf Q}}_{\beta)\mu\nu}{\sf S}^{\mu\nu}
+\f{1}{2}{{\sf Q}}_{\nu\mu(\alpha}{\sf S}^{\mu}_{\,\,\beta)}{\rm u}^\nu,
\eea
where the quantities $\rho$, $p_r$ and $p_t$ are the usual energy density, radial and tangential pressures of the fluid respectively, and ${\rm v}_{\mu}$ is a unit spacelike vector field in radial direction.
From the microscopical viewpoint, a randomly distributed gas of fermions is the source of spacetime torsion. However, we have to deal with this issue at a macroscopic level, which means that we need to perform a suitable spacetime averaging process. In this regard, the average of spin density tensor vanishes macroscopically, i.e., $\langle {\sf S}_{\mu\nu} \rangle=0$ \cite{ECT3}\cite{Gas} however, the square of spin density tensor ${\sf S}^2=\f{1}{2}\langle {\sf S}_{\mu\nu}{\sf S}^{\mu\nu}\rangle$ will survive to have contribution within the total {\sf EMT} \cite{Gas},\cite{Hehlgrg}
\bea\label{emtspgas}
{{\sf T}}_{\alpha\beta}^{{\rm total}}={{\sf T}}_{\mu\beta}+\theta_{\mu\beta}.
\eea
 Taking these considerations into account, the relations (\ref{ecfieldeq}) and (\ref{SSCI}) together with (\ref{FC}-\ref{emtspgas}) give the combined field equation with anisotropic matter distribution and spin correction terms as 
\be\label{EFESSP}
{\sf G}_{\mu\nu}-\Lambda g_{\mu\nu}=\kappa\left(\rho+p_t-\f{\kappa}{2}{\sf S}^2\right){\rm u}_{\mu}{\rm u}_{\nu}+\kappa\left(p_t-\f{\kappa}{4}{\sf S}^2\right)g_{\mu\nu}+(p_r-p_t){\rm v}_\mu {\rm v}_\nu.
\ee
We see that the contribution due to spin squared terms appear, effectively as a negative density, in the pressure profiles and energy density of the anisotropic fluid\footnote{In the present model, we assume the anisotropy of ordinary matter only within the {\sf EMT} of the fluid part and allow for a random distribution of the spinning particles. However, the existence of anisotropy within the spin part needs the study of a spin polarized matter distribution that can occur in the presence of a background magnetic field.}. In this sense, Hehl et al. have utilized the Weyssenhoff description of spinning fluid to show that such a contribution acts as a stiff matter \cite{ECT4,Gas}. Such a behavior plays an important role in super dense regimes of extreme gravity, even if the orientation of spinning particles is randomly distributed. This leads to gravitational repulsion and avoidance of curvature singularities by violating the energy condition of the singularity theorems \cite{ECT4}. Furthermore, it has been shown that the repulsive effects owing to the presence of spin effects would replace the big-bang singularity with a nonsingular big- bounce, before which the Universe was contracting \cite{spin-bounce}, \cite{spin-bounce1}. 
\par
\section{WORMHOLE GEOMETRIES}\label{WHS}
In this section we deal with dynamical wormhole solutions in the context of {\sf EC} gravity. The theoretical construction of wormhole spacetimes is usually performed by employing the method where, in order to obtain a desired metric, one is allowed to take the form of the metric functions freely, such as the redshift and shape functions, or even the scale factor for dynamical wormholes. We shall therefore follow the conventional method in order to find solutions in {\sf EC} gravity which are utilized also in theoretical cosmology. Moreover, we shall prescribe the matter content by specifying the {\sf {\sf EoS}} for the radial and tangential pressures and energy density and then we solve for the field equations to obtain the redshift and shape functions together with the scale factor.
\subsection{EVOLVING LORENTZIAN WORMHOLES}\label{WHS1}
We are interested in wormhole solutions in a cosmological background, thus, we generalize Morris and Thorne wormhole to a time-dependent spacetime given by
\begin{eqnarray}\label{evw}
ds^2= -e^{2\phi(r)}dt^2+R(t)^2 \left(
\frac{dr^2}{1-\frac{b(r)}{r}}+r^2(d\theta^2+\sin^2 \theta d
\varphi^2)\right),
\end{eqnarray}
where $R(t)$ is the scale factor of the Universe, $\phi(r)$ being the redshift function and $b(r)$ is the wormhole shape function. The shape function must satisfy the flare-out condition at the throat, i.e., we must have $b^{\prime}(r_0)<1$ and $b(r)<r$ for $r>r_0$ in the whole spacetime, where $r_0$ is the throat radius. In the present work, we consider $\phi(r)=0$ in order to ensure the absence of horizons and singularities throughout the spacetime. These evolving Lorentzian wormholes are conformally related to another family of static wormholes with zero-tidal force. The general constraints on these functions has been discussed by Morris and Thorne in~\cite{mt}. It is clear that if $b(r)$ and $\phi(r)$ tend to zero the metric~(\ref{evw}) becomes the flat {\sf FRW} metric, and as $R(t) \rightarrow constant$ the static Morris-Thorne wormhole is recovered. In the herein model, we search a way to determine the shape function $b(r)$ and the scale factor $R(t)$ in order to construct dynamical wormholes.
\par  
Let us now find the combined field equations for the metric (\ref{evw}). In an orthonormal reference frame, the nonzero components of the {\sf EMT} are given by
\bea
\rho(r,t)&=&3H^2+\frac{b^{\prime}(r)}{ R(t)^2r^2}+\frac{{\sf S}^2(t)}{4}\label{FE32}\\
p_r(r,t)&=&-3H^2-2\dot{H}-\frac{b^{\prime}(r)}{2 R(t)^2r^2}+\frac{b(r)}{2 R(t)^2r^3}+\frac{{\sf S}^2(t)}{4},\label{FEs11}\\
p_t(r,t)&=&-3H^2-2\dot{H}-\frac{b(r)}{R(t)^2r^3}+\frac{{\sf S}^2(t)}{4},\label{FE22}
\eea
where we have set the units so that $\kappa=1$ and use has been made of ${\rm u}_\mu=\left[1,0,0,0\right]$ and ${\rm v}_\mu=\left[0,\sqrt{1-b(r)/r},0,0\right]$ as the timelike and spacelike vector fields and $H=\frac{\dot{R}}{R}$. The conservation equation ${\sf T}^{\mu}_{\,\,\,\nu;\mu}=0$, leaves us with the following relations given as 
  \begin{eqnarray}\label{CE1}
  \dot{\rho}+H [3 \rho+p_r+2p_{t}]=\frac{1}{2}[\dot{{\sf S}}{\sf S}+3H{\sf S}^2], \\
  p^{\, \prime}_r+\frac{2(p_{t}-p_r)}{r}=0.
  \label{CE2}
  \end{eqnarray}
We note that the above equations describe energy conservation in the general case when all fields interact with each other. Let us consider the dependence of spin square density on the scale factor as ${\sf S}^2={\sf S}_{0}^2/a(t)^m$ ($m>0$)~\cite{spin-bounce1}. Also, from equation (\ref{CE2}) we see that for an isotropic pressure, i.e. $p_r=p_t$,  we have $p^{\, \prime}_r=0$, so the pressure will depend only on time which corresponds to the standard {\sf FRW} cosmologies. However, our aim here is to study evolving wormholes with anisotropic pressures in an inhomogeneous spacetime which merge smoothly to
the cosmological background. We therefore adopt the generalized {\sf {\sf EoS}} given by~\cite{geos}
\begin{eqnarray}
\rho=\frac{w}{1+2\gamma}(p_{r}+2\gamma p_t),
\label{e8}
\end{eqnarray}
where $w$ and $\gamma$ are constants and for $w=-1$ the above {\sf {\sf EoS}} reduces to a particular {\sf {\sf EoS}} already explored in \cite{ebriza}. Clearly, this equation reduces to a linear {\sf EoS} at spatial infinity where  $p_{r}=p_t$ . In order to obtain the two unknown functions, $b(r)$ and $R(t)$, we substitute equations (\ref{FE32})-(\ref{FE22}) into the above {\sf EoS} which yields
  \begin{eqnarray}\label{CE1}
  \frac{(1+\gamma(2+w))rb^{\prime}(r)-w(\gamma-1)b(r)}{(1+2\gamma)r^3 }=\frac{(1+2\gamma)R(t)^2[R(t)^{2m}(12H^2(w+1)+8w\dot{H})-{\sf S}_{0}^2(w-1)]}{(4+8\gamma)R(t)^{2m}}.
  \label{CE8}
  \end{eqnarray}
Fortunately, this equation can be separated into radial and temporal equations. Therefore, both sides can be set to constant value. Let us take the separation constant as ${\sf C}$ whence we get the shape function as
\begin{align}
b(r)=\frac{{\sf C} r^3}{3+w}+{\sf C_1} r^{\frac{w(\gamma-1)}{\gamma(2+w)+1}},\label{CE9}
\end{align}
where ${\sf C_1}$ is an integration constant. Using the condition $b(r_0)=r_0$ at the throat we have
\begin{align}
{\sf C_1}=\frac{r_0(-{\sf C}r_{0}^2+3+w)}{(3+w) r_0^{\frac{w(\gamma-1)}{\gamma(2+w)+1}}},
\end{align}
from which we find at the throat
\begin{align}
 b^{\prime}(r_0)=\frac{{\sf C}r_{0}^2(2\gamma+1)+w(\gamma-1))}{\gamma(2+w)+1},
\label{cebp}
\end{align}
where we have used Eq. (\ref{CE9}).  From equation (\ref{CE8}) we obtain the following master equation for the scale factor
\begin{align}
\bigg[\big(3(1+w)H^2+2w\dot{H}\big)R(t)^{2m}-\frac{(w-1){\sf S}_0^2}{4}\bigg]R(t)^2+{\sf C}R(t)^{2m}=0.
\label{hht}
\end{align}
Note that the rate of expansion of these evolving wormholes is only determined by the {\sf EoS} parameter $w$ and is independent of parameter $\gamma$. Integrating the above equation leaves us with the following equation as
 \begin{align}
3H(t)^2+\frac{3{\sf C}}{R(t)^2}=kR(t)^{-3(1+w)}+\frac{3{\sf S}_0^2(w-1)R(t)^{-2m}}{4(2m-3(1+w))}.
\label{exp2}
 \end{align}
This differential equation is the standard Friedmann equation in the presence of torsion~\cite{ziaieepjc}.
In the following subsections, with the help of the master equation, we will determine the behavior of scale factor and the related properties of the energy conditions within the wormhole geometry. Thus, in order to study an evolving wormhole in detail, we consider two cases ${\sf C}=0$ and ${\sf C}\neq0$.
\subsection{Solutions for the case ${\sf C}=0$}\label{3.2ew}
Setting the ${\sf C}=0$ in Eq. (\ref{CE9}), we find the shape function as 
\bea\label{shapec0}
b(r)=r_0\left(\frac{r_0}{r}\right)^{\frac{w(1-\gamma)}{1+\gamma(2+w)}}.
\eea
Also the  condition $ b^{\prime}(r_0)<1$ in equation (\ref{cebp}) gives the following conditions
\bea\label{condswg}
-2\gamma-1<w~~\textrm{or}~~ w<-\frac{1+2\gamma}{\gamma}~~~~\textrm{for}~~\gamma>0,\nn
-\frac{1+2\gamma}{\gamma}<w<-2\gamma-1~~~~\textrm{for}~~~\gamma<-\frac{1}{2},\nn
-2\gamma-1<w<-\frac{1+2\gamma}{\gamma}~~~~~\textrm{for}~~~~-\frac{1}{2}<\gamma<0.
\eea
We note that these wormhole solutions are asymptotically flat. Moreover, in order to have a wormhole spacetime for $\gamma=0$ and $R(t)=constant$, it is clear that we must have $w>0$ or $w<-1$. These static wormhole solutions were firstly considered in~\cite{lsf}. Next, we proceed find the scale factor for this solution.
 \subsubsection{Specific case: $w>0$} 
Let us now investigate the features of the evolving wormhole. We firstly solve the differential equation (\ref{hht}) to find the scale factor for {\sf {\sf GR}} case (${\sf S}_0=0$) as
 \begin{align}
 R(t)=({\sf C_2}+{\sf C_3}t)^{\frac{2w}{3(1+w)}},
\end{align}
where ${\sf C_2}$ and ${\sf C_3}$ are integration constants. Using the field equations (\ref{FE32})-(\ref{FE22}) we obtain
\begin{align}
\rho(r,t)=\frac{4{\sf C_3}^2w^2}{3(1+w)^2({\sf C_2}+{\sf C_3}t)^2}+\frac{wr_0^{-2}\big(\gamma-1\big)}{({\sf C_2}+{\sf C_3}t)^{\frac{4w}{3(1+w)}}\big(1+\gamma(2+w)\big)}\Bigg[\frac{r_0}{r}\Bigg]^{\frac{(3+w)(1+2\gamma)}{1+\gamma(2+w)}},
\label{wpo1}
\end{align}
\begin{align}
\rho(r,t)+p_r(r,t)=\frac{4{\sf C_3}^2w}{3(1+w)({\sf C_2}+{\sf C_3}t)^2}-\frac{r_0^{-2}\big(1+2\gamma+w\big)}{({\sf C_2}+{\sf C_3}t)^{\frac{4w}{3(1+w)}}\big(1+\gamma(2+w)\big)}\Bigg[\frac{r_0}{r}\Bigg]^{\frac{(3+w)(1+2\gamma)}{1+\gamma(2+w)}},
\label{wpo2}
\end{align}
and 
\begin{align}
\rho(r,t)+p_t(r,t)=\frac{4{\sf C_3}^2w}{3(1+w)({\sf C_2}+{\sf C_3}t)^2}+\frac{r_0^{-2}\big(1+2\gamma+w(2\gamma-1)\big)}{2({\sf C_2}+{\sf C_3}t)^{\frac{4w}{3(1+w)}}\big(1+\gamma(2+w)\big)}\Bigg[\frac{r_0}{r}\Bigg]^{\frac{(3+w)(1+2\gamma)}{1+\gamma(2+w)}}.
\label{wpo3}
\end{align}
Notice that for reasonable values of $\gamma$, the components of $\rho$, $\rho+p_r$ and $\rho+p_t$ tend to zero as $t\rightarrow\infty$. We can also suitably choose the constants so that the {\sf WEC} be satisfied at the wormhole throat. Moreover, we see that for $w>0$ we obtain $\frac{4w}{3(1+w)}<2$. Hence, at late times, the second term of Eqs. (\ref{wpo1})-(\ref{wpo3}) would determine the sign of these equations at infinity. Thus, in order to fulfill the {\sf {\sf WEC}} we must have $w>-\frac{1+2\gamma}{\gamma}$ for $-\frac{1}{2}<\gamma<0$ and  $-1-2\gamma<w$ for $\gamma<-\frac{1}{2}$ . In this case, these conditions violate the flaring-out condition at the throat ($b^{\prime}(r_0)<1$) and consequently the {\sf {\sf WEC}} is violated at late times.
\par
\begin{figure}
\begin{center}
\includegraphics[scale=0.26]{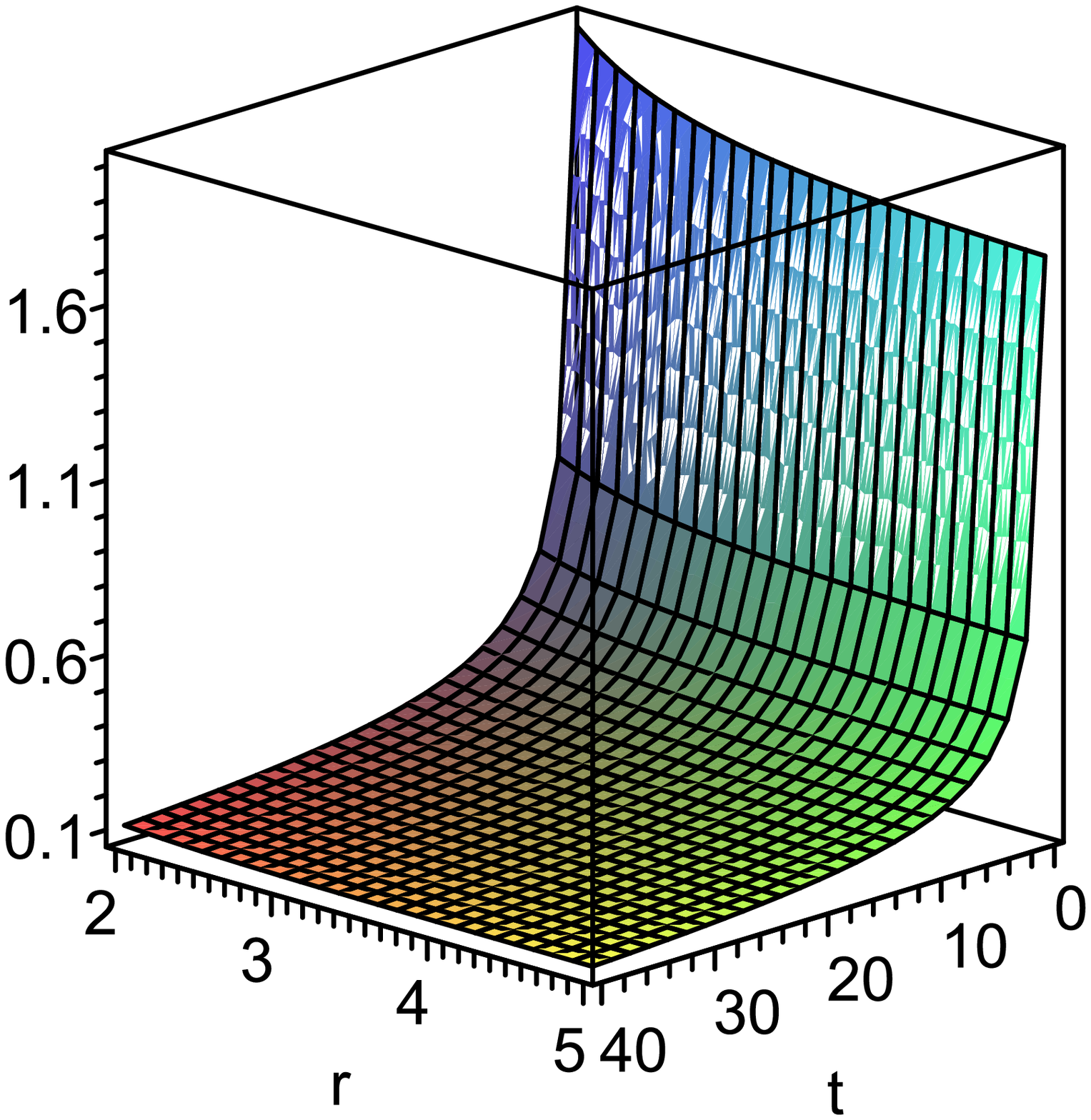}
\includegraphics[scale=0.26]{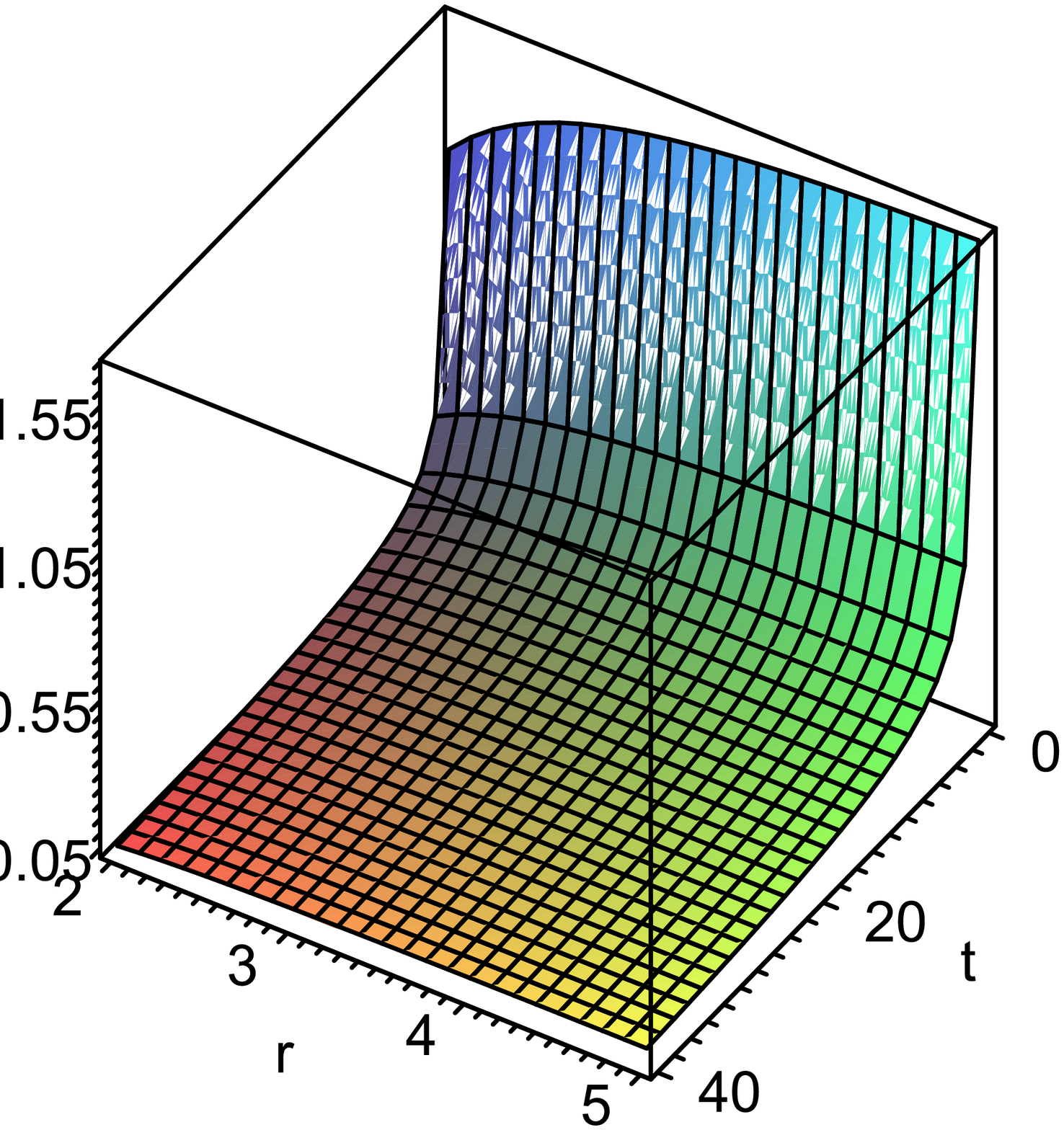}
\includegraphics[scale=0.26]{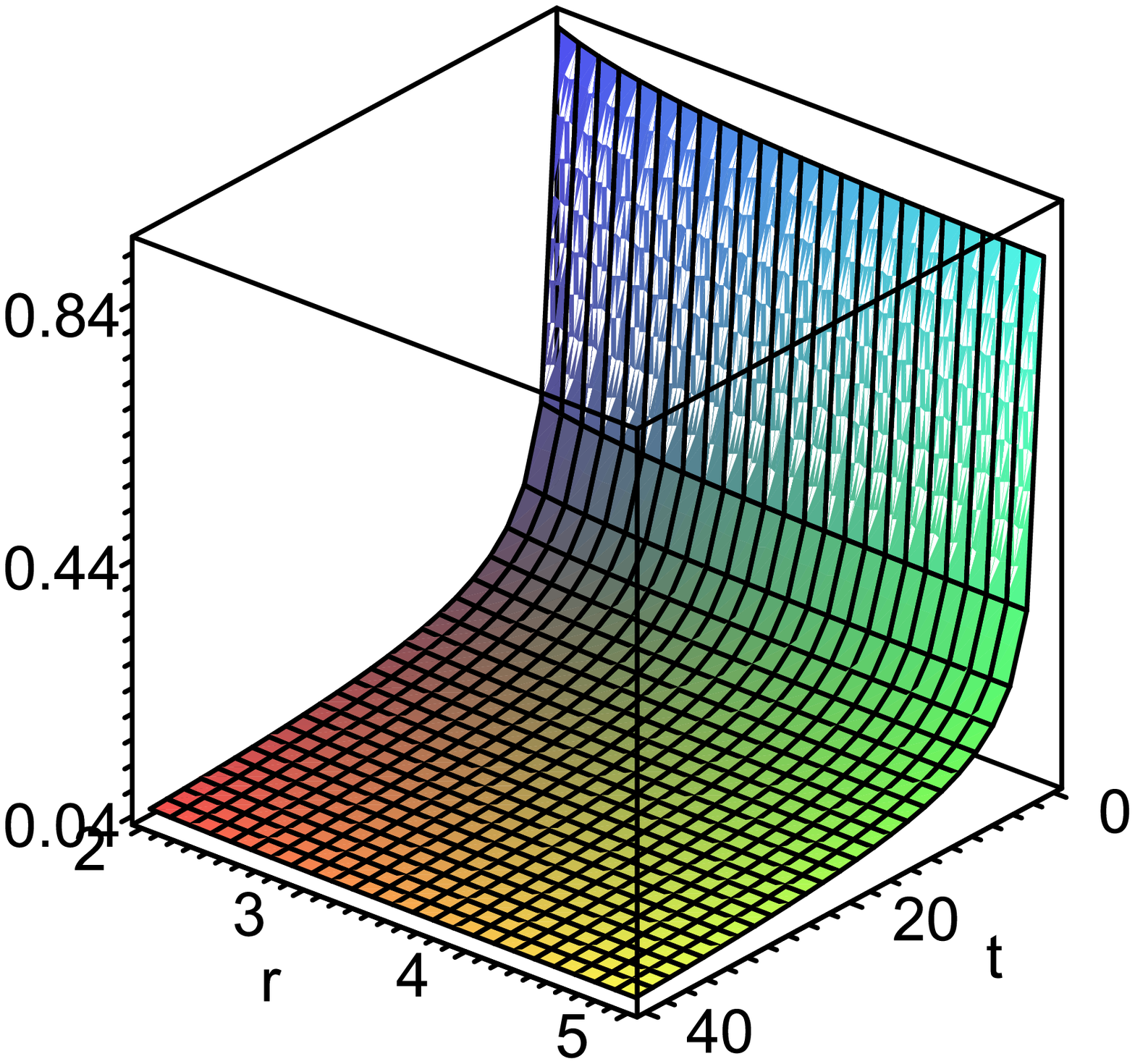}
\caption {The behavior of $\protect\rho $ , $\protect\rho +p_{r}$  and $\protect\rho +p_{t}$  versus $r$ respectively from left to right, for $w=1$, $m=1$, $\gamma=2$, ${\sf C_4}={\sf C_5}=0.3$ , $r_0=2$ and ${\sf S}_0=1$.}\label{fg1}
\end{center}
\end{figure}

 Now we solve the differential equation (\ref{hht}) for ${\sf S}_0\neq0$. Since solving the differential equation (\ref{hht}) is too
 complicated, in general, we will consider restrictions on the state parameters $w$ and $m$ and then solve Eq. (\ref{hht}) to obtain the scale factor. Firstly, let  us consider a stiff matter with the {\sf {\sf EoS}} $w=1$ in equation (\ref{hht}). Then, the scale factor is found as
\begin{align}
R(t)=({\sf C_4}+{\sf C_5}t)^{\frac{1}{3}},
\end{align}
whereby we get the following expressions as
 \begin{align}
 \rho(r,t)=\frac{{\sf C_5}^2}{3({\sf C_4}+{\sf C_5}t)^2}+\frac{r_0^{-2}\big(\gamma-1\big)}{({\sf C_4}+{\sf C_5}t)^{\frac{2}{3}}\big(1+3\gamma\big)}\Bigg[\frac{r_0}{r}\Bigg]^{\frac{4(1+2\gamma)}{1+3\gamma}}+\frac{{\sf S}_0^2}{4({\sf C_4}+{\sf C_5}t)^{\frac{2m}{3}}},
 \end{align}
 \begin{align}
 \rho(r,t)+p_r(r,t)=\frac{2{\sf C_5}^2}{3({\sf C_4}+{\sf C_5}t)^2}-\frac{r_0^{-2}\big(\gamma+1\big)}{({\sf C_4}+{\sf C_5}t)^{\frac{2}{3}}\big(1+3\gamma\big)}\Bigg[\frac{r_0}{r}\Bigg]^{\frac{4(1+2\gamma)}{1+3\gamma}}+\frac{{\sf S}_0^2}{2({\sf C_4}+{\sf C_5}t)^{\frac{2m}{3}}},
 \end{align}
 and
 \begin{align}
 \rho(r,t)+p_t(r,t)=\frac{2{\sf C_5}^2}{3({\sf C_4}+{\sf C_5}t)^2}+\frac{r_0^{-2}\big(2\gamma\big)}{({\sf C_4}+{\sf C_5}t)^{\frac{2}{3}}\big(1+3\gamma\big)}\Bigg[\frac{r_0}{r}\Bigg]^{\frac{4(1+2\gamma)}{1+3\gamma}}+\frac{{\sf S}_0^2}{2({\sf C_4}+{\sf C_5}t)^{\frac{2m}{3}}},
\end{align}
where ${\sf C_4}$ and ${\sf C_5}$ are integration constants. It is Noteworthy that if $-\frac{1}{3}<\gamma$ or $\gamma<-1$ we have asymptotically flat evolving wormholes for which the flare-out condition is satisfied. In this case, in order that the {\sf {\sf WEC}} be satisfied at late times, we find that for $m\leq1$, it is the third term in above expressions that decides the sign of them. Hence, by making a suitable choice for ${\sf S}_0$ the {\sf {\sf WEC}} can be satisfied. From another side, we see that for $m>1$ the sign of second term within these expressions is positive for $-1<\gamma<-\frac{1}{3}$ at late times which leads to the fulfillment of {\sf {\sf WEC}}; however, the flare-out condition is violated. Fig.(\ref{fg1}) shows that it is possible to choose suitable values for the constants in order to satisfy the {\sf {\sf WEC}} in whole spacetime.
 
\subsection{Solutions for the case ${\sf C} \neq 0$}\label{3.3ew}
\begin{figure}
\begin{center}
\includegraphics[scale=0.4]{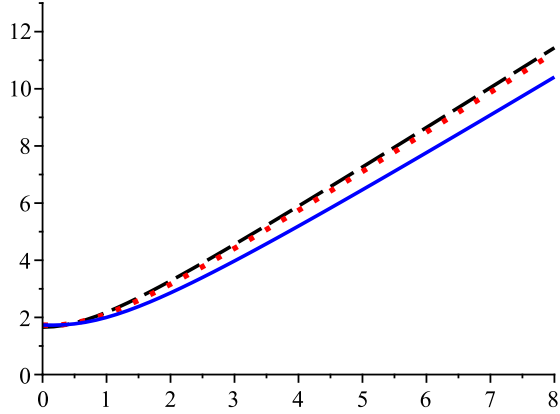}
\caption{The phase space diagram $\dot{R}$ versus $R$ for $k=2$, ${\sf C}=-3$, ${\sf S}_0=1$ and $m=0.25, 0.5, 1$.}\label{fg2}
\end{center}
\end{figure}

Let us proceed with the general case in which ${\sf C}\neq0$. An interesting case is that of traversable wormholes
supported by the dark energy {\sf {\sf EoS}} ($-1 \leq w < -1/3$) that is required for cosmic acceleration. For instance, let
us consider $w=-1$. Then, Eq. (\ref{exp2}) leads to
\begin{align}
\dot{R}^2=\frac{kR(t)^2}{3}-\frac{{\sf S_0}^2}{4mR(t)^{2(m-1)}}-{\sf C}
\label{exp1}
\end{align}
where ${k}$ is an integration constant. Since this equation cannot be integrated analytically for $R(t)$, it would be instructive to plot the phase space diagram of $R$ versus $R(t)$. Fig. (\ref{fg2}) shows the  phase space constructed out of the scale factor and its time derivative for a few values of the parameter $m$. As is seen for large values of the scale factor, $\dot{R}$ depends linearly on $R$ and for ${\sf S}_0=0$ and ${\sf C}=0$ in equation (\ref{exp1}) we have an exponential expansion for an inflating wormhole i.e, $R(t)=R_0e^{\sqrt{\frac{k}{3}}t}$.
\par
For $m=1$ within the master Eq.(\ref{hht}), the scale factor is found as
\begin{align}
R(t)=\frac{{\sf S_1}+{\sf C_6}^{\!\!2} e^{2{\sf C_7}t}}{4{\sf C_6}{\sf C_7}e^{{\sf C_7}t}},
\label{ee2}
\end{align}
where ${\sf C_6}$ and ${\sf C_7}$ are integration constants and ${\sf S_1}=4{\sf C}+{\sf S_0}^{\!\!2}$. In case we set ${\sf S_1}=0$ and ${\sf C_7}>0$, which gives an exponential expansion for an inflating wormhole i.e, $R(t)=R_0e^{{\sf C_7}t}$, the wormhole solution presented in \cite{ebriza} is recovered. We therefore note that without loss of generality, by rescaling the  constants ${\sf S}_1>0$ and ${\sf S}_1<0$, the scale factor can be found as $R(t)=R_0\sinh({\sf C_7}t)$ and   $R(t)=R_0\cosh({\sf C_7}t)$ respectively. 
\par
\begin{figure}
   \begin{center}
   \includegraphics[scale=0.27]{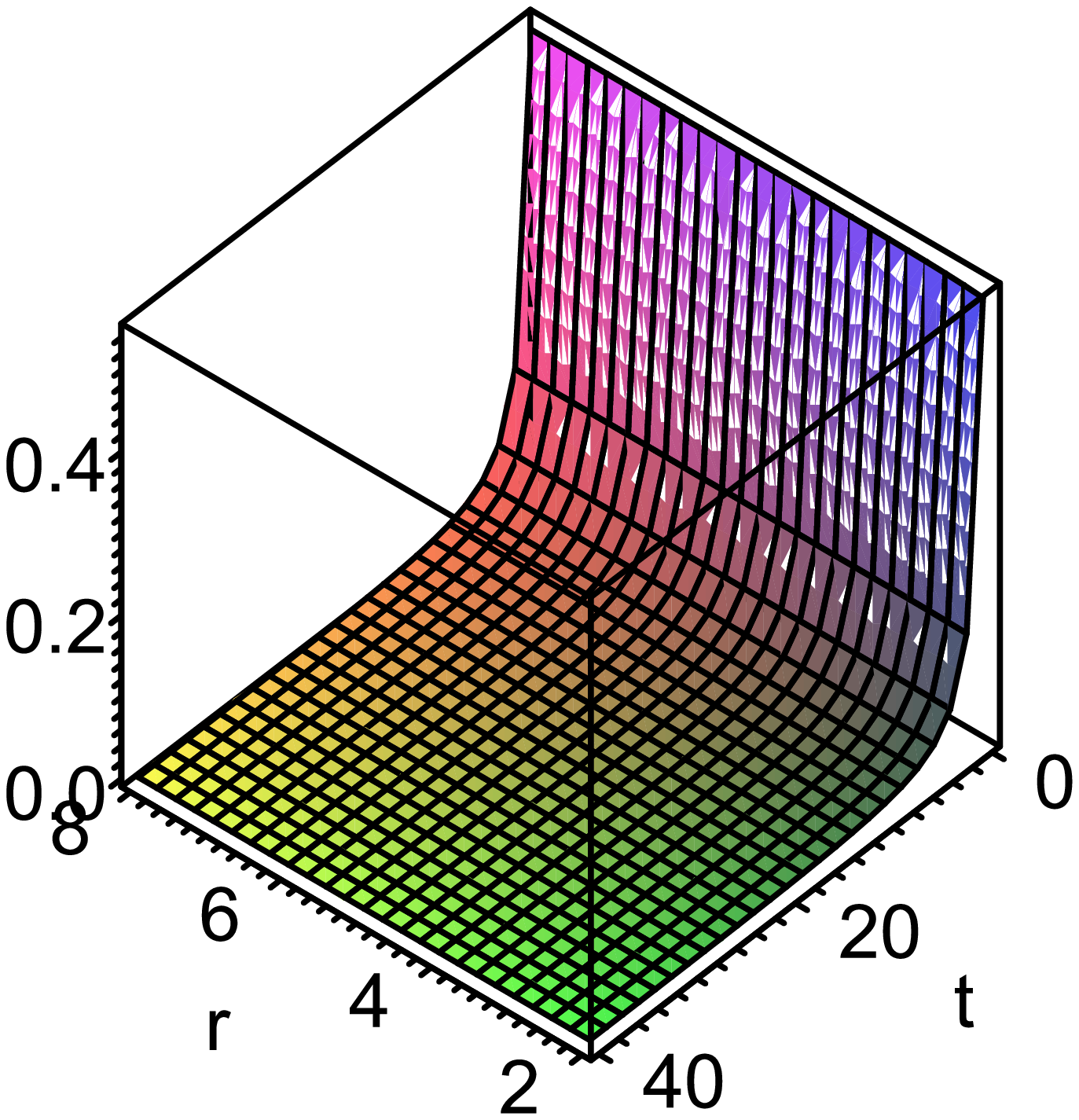}
   \includegraphics[scale=0.23]{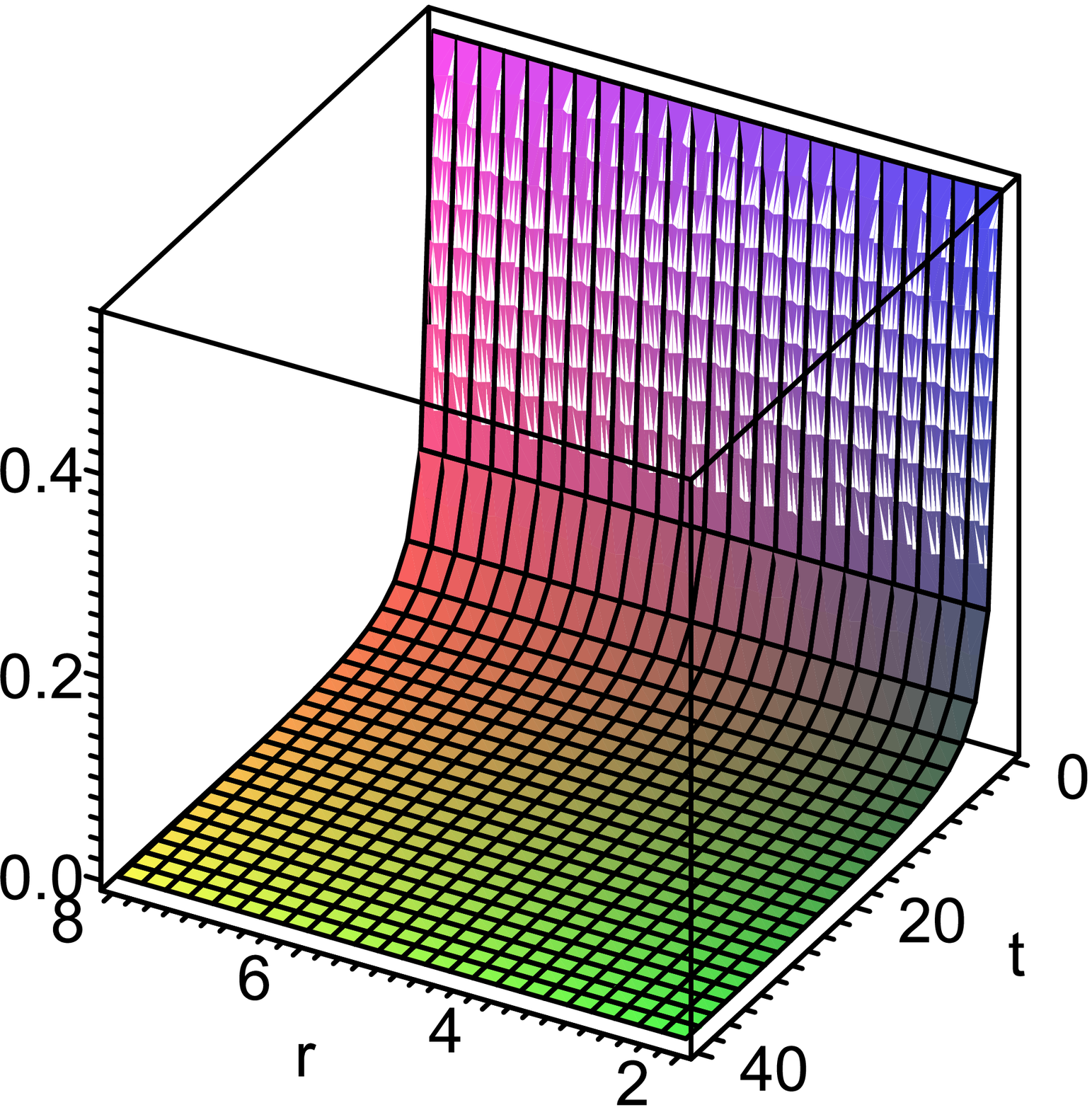}
   \includegraphics[scale=0.27]{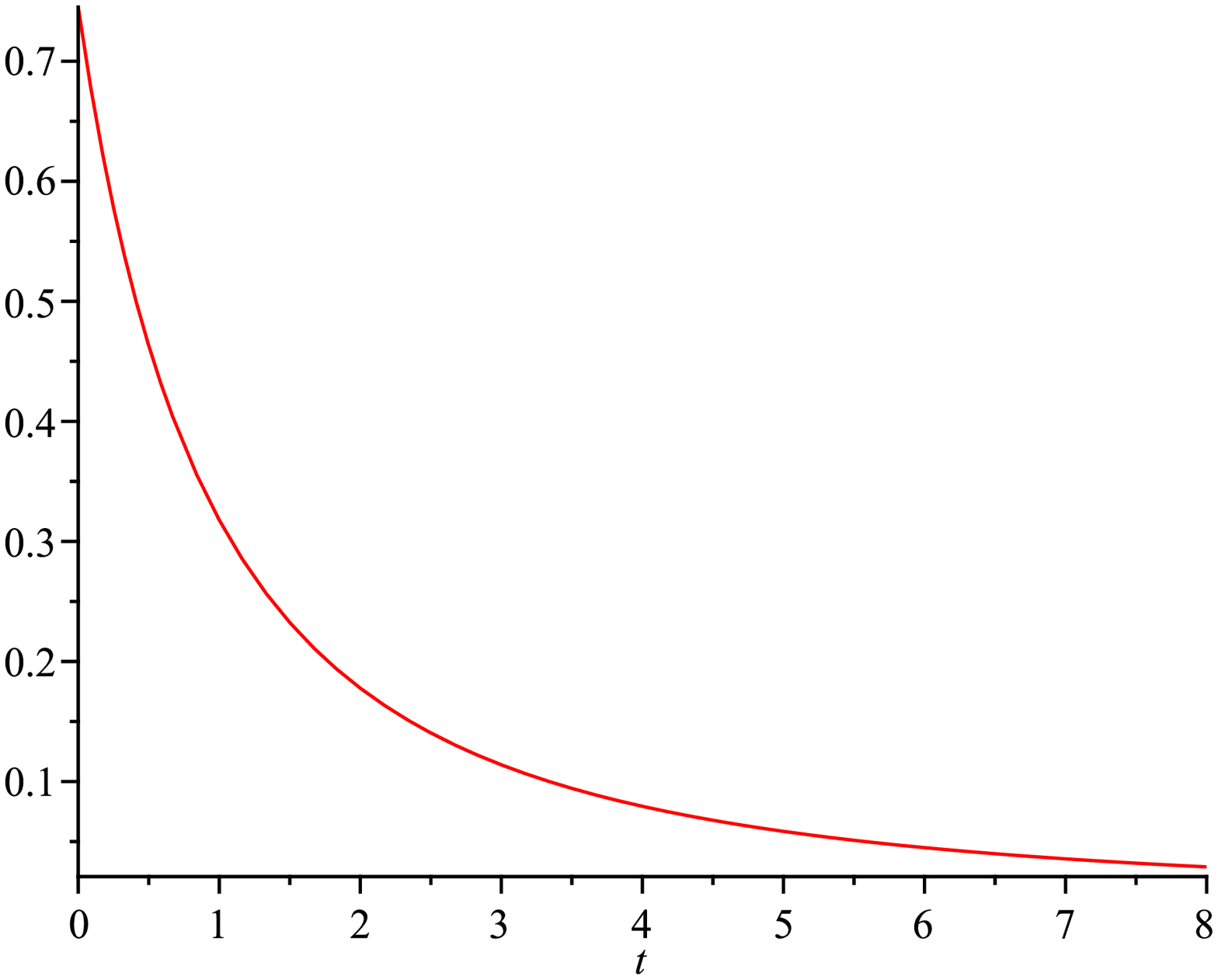}
   \caption {The behavior of $\protect\rho $ , $\protect\rho +p_{r}$  and $\protect\rho +p_{t}$  versus $r$ respectively from left to right  for $w=3$, $\gamma=1$, ${\sf C}=-0.2$, $r_0=2$, $\beta_1=16.83$, $\beta_2=72$, $\beta_3=72$ and ${\sf {\sf S}_0}=1$.}\label{fg3}
   \end{center}
   \end{figure}

Using the field equations for scale factor (\ref{ee2}) we obtain
\begin{align}
\rho(r,t)=\frac{\big[2{\sf C_6}^{\!\!2}(12{\sf C}-{\sf S_0}^{\!\!2})e^{2{\sf C_7}t}+3{\sf C_6}^{\!\!4}e^{4{\sf C_7}t}+3{\sf S_1}^{\!\!2}\big]{\sf C_7}^{\!\!2}}{({\sf S_1}+{\sf C_6}^{\!\!2}e^{2{\sf C_7}t})^2}+\frac{16{\sf C_6}^{\!\!2}{\sf C_7}^{\!\!2}(\gamma-1)({\sf C}r_0^2-1)}{(\gamma+1)r_0^2({\sf S_1}+{\sf C_6}^{\!\!2}e^{2{\sf C_7}t})^2}\bigg[{\frac{r_0}{r}}\bigg]^{\frac{2(2\gamma+1)}{\gamma+1}},
\end{align}
 \begin{align}
 \rho(r,t)+p_r(r,t)=\frac{32\gamma {\sf C_6}^{\!\!2}{\sf C_7}^{\!\!2}({\sf C}r_0^2-1)e^{2{\sf C_7}t}}{(\gamma+1)r_0^2({\sf S_1}+{\sf C_6}^{\!\!2}e^{2{\sf C_7}t})^2}\bigg[{\frac{r_0}{r}}\bigg]^{\frac{2(2\gamma+1)}{\gamma+1}},
 \end{align}
 and
 \begin{align}
 \rho(r,t)+p_t(r,t)=-\frac{16 {\sf C_6}^{\!\!2}{\sf C_7}^2({\sf C}r_0^2-1)e^{2{\sf C_7}t}}{(\gamma+1)r_0^2({\sf S_1}+{\sf C_6}^{\!\!2}e^{2{\sf C_7}t})^2}\bigg[{\frac{r_0}{r}}\bigg]^{\frac{2(2\gamma+1)}{\gamma+1}}.
 \end{align}
 Now, in order to check the {\sf WEC}, we first investigate the
 behavior of $\rho+p_r(r_0)$ and $\rho+p_t(r_0)$ at time $t=0$, which is given by
  \begin{align}
  \rho(r_0)+p_r(r_0)=\frac{32{\sf C_6}^2{\sf C_7}^2({\sf C}r_0^2-1)\gamma}{r_0^2({\sf S_1}+{\sf C_6}^2)^2(\gamma+1)},
  \end{align}
  and
 \begin{align}
 \rho(r_0)+p_t(r_0)=-\frac{16{\sf C_6}^2{\sf C_7}^2({\sf C}r_0^2-1)}{r_0^2({\sf S_1}+{\sf C_6}^2)^2(\gamma+1)}.
 \end{align} 
We therefore observe that if we set ${\sf C}=0$ in the above expressions, the conditions $ \rho(r_0)+p_r(r_0)>0$ and $\rho(r_0)+p_t(r_0)>0$ will be satisfied when  $-1<\gamma<0$. However such a restriction on $\gamma$ parameter is in contradiction with condition $b^{\prime}(r_0)<1$ which imposes $0<\gamma$ or $\gamma<-1$. Thus, the {\sf {\sf WEC}} (and also {\sf {\sf NEC}}) is violated in the vicinity of the wormhole throat. For ${\sf C}\neq0$ and ${\sf C}r_0^2-1<0$, if $-1<\gamma<0$ the {\sf {\sf NEC}} is satisfied at the wormhole throat. Also, if $-\frac{1}{2}<\gamma<0$  the condition $b^{\prime}(r_0)<1$ will be satisfied. Hence, one can choose $-\frac{1}{2}<\gamma<0$ in order that the {\sf {\sf WEC}} be satisfied at the wormhole throat. For ${\sf C}r_0^2-1>0$ if $\gamma<-1$ the {\sf {\sf NEC}} is satisfied at the wormhole throat, but the flare-out condition ($-1<\gamma<-\frac{1}{2}$) is violated. We further note that for an inflating wormhole, (${\sf C_7}>0$) the values $\rho$, $p_r$ and $p_t$ for large $r$ will take the following form as
\begin{align}
\rho=-p_r=-p_t=3{\sf C_7}^2,
\end{align}
which corresponds to dark energy {\sf EoS} (cosmological constant). Also, it is clear that both $\rho+p_r$ and $\rho+p_t$ tend to zero as $r$ tends to infinity. Since these two expressions have no real positive root, in order to find their sign, it is
sufficient to investigate the sign at the throat. Therefore, in order that the {\sf {\sf WEC}} be satisfied throughout the spacetime, one can impose the condition on satisfaction of {\sf {\sf WEC}} at wormhole throat. Finally, in order to investigate the traceless {\sf {\sf EMT}} case, we set $w=3$ and $\gamma=1$ in equation (\ref{CE9}). Applying these choices, the shape function is found as\begin{align}
b(r)={\sf C}(r^3-r_0^3)+r_0,
\end{align}
and the condition $b^{\prime}(r_0)=3{\sf Cr_0}^2<1$ must satisfied at the throat. Now, from Eq (\ref{exp2}), we find the scale factor for $m=1$ as
\begin{align}
R(t)=\big[\beta_1t^2+\beta_2t+\beta_3\big]^{\frac{1}{2}},
\end{align}
where $\beta_1={\sf S_0}^2/12-{\sf C}$ and $\beta_2$ and $\beta_3$ are constants of integration. Notice that for expanding wormholes, we must have $\beta_1>0$ and $\beta_2-4\beta_1\beta_3<0$. In order to check the fulfillment of {\sf {\sf WEC}} we need to study the behavior of $\rho$, $\rho+p_r$ and $\rho+p_t$ for large $t$, given by
 \begin{align}
 \rho(r,t)=\frac{{\sf S_0}^2}{6\beta_1t^2}+{\mathcal O}\left(\frac{1}{t^3}\right),
 \end{align}
 \begin{align}
 \rho(r,t)+p_r(r,t)=\frac{(2{\sf S_0}^2r^3+3{\sf C}r_0^3-3r_0)}{144\beta_1r^3t^2}+{\mathcal O}\left(\frac{1}{r^3t^3}\right),\label{tr1}
 \end{align}
 and
  \begin{align}
  \rho(r,t)+p_t(r,t)=\frac{(4{\sf S_0}^2r^3-3{\sf C}r_0^3+3r_0)}{288\beta_1r^3t^2}+{\mathcal O}\left(\frac{1}{r^3t^3}\right).
  \label{tr2}
  \end{align}
It is seen that for $\beta_1>0$ and $-\frac{2{\sf S_0}^2r_0^3-3}{3r_0^2}<{\sf C}\leq 0$ , both of $\rho+p_r$ and $\rho+p_t$ are positive at wormhole throat. Also, the value of energy density is positive for $\beta_1>0$.  It can be shown that the {\sf {\sf WEC}} for $t=0$ is satisfied at the throat, if we choose suitable values for the constant parameters $\beta_2$, $\beta_3$ and ${\sf S_0}$. We further note that in {\sf {\sf GR}} where ${\sf S_0}=0$, the {\sf {\sf NEC}} is violated for large time as is seen from equations (\ref{tr1})-(\ref{tr2}). However, the {\sf WEC} can be satisfied as is shown in  Fig. (\ref{fg3}).

\section{Concluding Remarks}\label{concluding}
In this paper we investigated evolving wormhole solutions in an expanding Universe in the framework of {\sf ECT} by considering  a Weyssenhoff spinning fluid along with an anisotropic energy momentum tensor ({\sf EMT}) for matter fields as supporting material for wormhole geometry. Assuming a specific form for the {\sf EoS} of perfect fluid part of the {\sf EMT}, the solutions are found using the method of separation of variables along with assuming a constant redshift function and a generalized form for the {\sf FRW} metric. It is interesting to note that the time evolution of these solutions is governed through an equation for the standard Friedmann equation with non-vanishing contribution for the spin effects. Two classes of solutions were obtained for zero and nonzero separation constant and the scale factor were calculated for these solutions. It was found that in {\sf GR}, these class of solutions do not satisfy {\sf WEC} throughout the spacetime, however, for a suitable choose of parameters in {\sf ECT} (${\sf S_0}$ and $m$), the energy conditions can be satisfied for all solutions and for whole spacetime. It is also worth mentioning that these solutions can be compared to evolving Lorentzian wormholes with the radial tension and tangential pressure having barotropic {\sf EoS} for which the {\sf WEC} is always violated~\cite{cspp}. Another important issue that needs to be investigated is the stability of solutions under small perturbations. As we know, the study of stability conditions of any static or stationary energy-momentum distribution under small perturbations is mandatory for its steady existence in the Universe. Traversable Lorentzian wormholes which have attracted a great deal of attention in recent modern research in {\sf GR} and its alternative theories are not an exception since a wormhole geometry which is of physical interest should survive enough so that its traversability is sensible. Therefore the stability of a given wormhole configuration becomes a central aspect of its study and in this regard the stability issue of wormholes has been investigated for the case of small perturbations retaining the original symmetry of the configurations. In particular, Poisson and Visser \cite{pv} have developed a straightforward method in order to analyze the stability features of thin-shell wormholes. This method has been widely utilized to investigate the stability aspects of more general spherically symmetric configurations~\cite{stabfeatureconf}. Work along this line has been pursued over the past years and much effort has been devoted to examine the stability of wormhole solutions in {\sf GR}~\cite{stabilityrefs1} and in some generalized theories of gravity~see e.g.,\cite{thi} and \cite{stablegeneralizedgr}. Concerning the above discussions, the stability analysis of wormhole solutions obtained here under small perturbations is an important problem, however, this task is beyond the scope of the present work and we postpone a detailed discussion on this issue to future studies.

\end{document}